\documentclass[10pt,conference]{IEEEtran}
\usepackage{cite}
\IEEEoverridecommandlockouts
\usepackage{amsmath,amssymb,amsfonts}
\usepackage{algorithmic}
\usepackage{graphicx}
\usepackage{textcomp}
\usepackage{xcolor}
\usepackage{silence}
\usepackage{soul}
\usepackage{xspace}
\usepackage{enumitem}
\usepackage{hyperref}
\usepackage{booktabs}
\usepackage{cleveref}
\usepackage{listings} 
\usepackage{graphicx}
\usepackage{subcaption}
\def\BibTeX{{\rm B\kern-.05em{\sc i\kern-.025em b}\kern-.08em
    T\kern-.1667em\lower.7ex\hbox{E}\kern-.125emX}}

\definecolor{ghostwhite}{rgb}{0.97, 0.97, 1.0} 
\usepackage{tikz}

\newcommand{\para}[1]{%
  \vspace{0.1 cm}%
  \noindent\textbf{#1.\ }%
}

\crefformat{section}{§#2#1#3}
\Crefformat{section}{§#2#1#3}

\newcommand*{\eg}{e.g.\@\xspace}

\newcommand{\tool}{\texttt{ReFuzzer}\xspace} 
\newcommand{\white}{\texttt{WhiteFox}\xspace} 
\newcommand{\grey}{\texttt{Fuzz4All}\xspace}
\newcommand{\black}{\texttt{BlackBox}\xspace} 
\newcommand{\ollama}{\texttt{Ollama}\xspace} 
\def\BibTeX{{\rm B\kern-.05em{\sc i\kern-.025em b}\kern-.08em
    T\kern-.1667em\lower.7ex\hbox{E}\kern-.125emX}}

\usepackage{multirow}
\usepackage{balance}
\usepackage{fancyvrb}
\usepackage{tcolorbox}
\usepackage{colortbl}
\usepackage{url}
\usepackage{listings}
\usepackage{cleveref}
\DeclareUnicodeCharacter{2060}{}

\newtheorem{example}{Example}

\newcommand{\code}[1]{\texttt{#1}}
\newcommand{\scode}[1]{{\scriptsize{\texttt{#1}}}}  

\crefformat{section}{§#2#1#3}
\Crefformat{section}{§#2#1#3}

\usepackage[T1]{fontenc}
\definecolor{mGreen}{rgb}{0,0.6,0}
\definecolor{mGray}{rgb}{0.5,0.5,0.5}
\definecolor{mPurple}{rgb}{0.58,0,0.82}
\definecolor{backgroundColour}{rgb}{0.99,0.99,0.97}
\definecolor{white}{rgb}{1.0, 1.0, 1.0}
\definecolor{ghostwhite}{rgb}{0.97, 0.97, 1.0}
\lstdefinestyle{CStyle}{
    backgroundcolor=\color{backgroundColour},   
    commentstyle=\color{mGreen},
    keywordstyle=\color{magenta},
    numberstyle=\color{mGray},
    stringstyle=\color{mPurple},
    basicstyle=\tt\fontsize{5.5}{5.5}\selectfont,
    breakatwhitespace=false,         
    breaklines=true,                 
    captionpos=b,
    keepspaces=true,                 
    numbers=left,
    stepnumber=1,
    firstnumber=1,
    numbersep=6pt,                  
    showspaces=false,                
    showstringspaces=false,
    showtabs=false,                  
    tabsize=2,
    language=C
}

\newtcolorbox[use counter=prompt]{promptbox}[2][]{%
  colback=backgroundColour,
  colframe=black,
  boxrule=0.7pt,
  before skip=5pt,
  after skip=5pt,
  left=2pt,
  right=2pt,
  top=2pt,
  bottom=2pt,
  fonttitle=\bfseries,
  title={Prompt~\theprompt: #2},
  label={#1}
}

\let\oldthebibliography\thebibliography
\let\endoldthebibliography\endthebibliography
\renewenvironment{thebibliography}[1]{%
  \oldthebibliography{#1}%
  \setlength{\itemsep}{0pt}%
  \setlength{\parskip}{0pt}%
}%
{\endoldthebibliography}

\begin{document}
\bstctlcite{IEEEexample:BSTcontrol}
\title{ReFuzzer: Feedback-Driven Approach to Enhance Validity of LLM-Generated Test Programs}

\author{\IEEEauthorblockN{1\textsuperscript{st} Iti Shree}
\IEEEauthorblockA{\textit{King's College London}\\
London, United Kingdom\\
iti.shree@kcl.ac.uk}
\and
\IEEEauthorblockN{2\textsuperscript{nd} Karine Even-Mendoza}
\IEEEauthorblockA{\textit{King's College London} \\
London, United Kingdom \\
karine.even\_mendoza@kcl.ac.uk}
\and
\IEEEauthorblockN{3\textsuperscript{rd} Tomasz Radzik}
\IEEEauthorblockA{\textit{King's College London} \\
London, United Kingdom \\
tomasz.radzik@kcl.ac.uk}
}
\maketitle

\begin{abstract}  

Existing LLM-based compiler fuzzers often produce syntactically or semantically invalid test programs, limiting their effectiveness in exercising compiler optimisations and backend components. We introduce \textbf{\tool{}}, a framework for refining LLM-generated test programs by systematically detecting and correcting compilation and runtime violations (\eg{} division by zero or array out-of-bounds accesses). \tool{} employs a feedback loop with a local LLM to validate and filter erroneous programs before execution, improving fuzzing effectiveness beyond crash detection and enabling the generation of diverse yet valid test programs.

We evaluated \tool{}'s effectiveness across black-, grey- and white-box fuzzing approaches targeting LLVM\-/Clang. \tool{} improved test programs' validity from 47.0–49.4\% to 96.6–97.3\%, with an average processing time of 2.9–3.5 s per test program on a dual-GPU machine. Further, refuzzing significantly increased code coverage in critical optimisation and IR generation components. For example, \textit{vectorization} coverage had an absolute of 9.2\%, 2.3\%, and 7.1\% improvement in black-, grey-, and white-box fuzzing, enhancing testing effectiveness.
\end{abstract}  


\section{Introduction}
\label{sec:intro}
\textit{Compiler fuzzing} has uncovered thousands of bugs, leading to fixes and significantly contributing to compiler reliability over the past decade \cite{c3:Yang2011,EMI2014,even2023grayc,c5:Xia2024}. The emergence of \textit{large language models} (LLMs) in code generation has made them valuable for compiler fuzzing \cite{c4:Yang2023,c5:Xia2024} due to their language-agnostic nature and ability to generate diverse programs. Such test programs are primarily used to detect compiler hangs and crashes (where the compiler fails to complete due to internal errors). It is because LLMs often struggle to produce valid code, reporting a 23.02\%–49.05\% static validity rate during fuzzing~\cite{c5:Xia2024}. 

We follow the terminology of \texttt{GrayC}~\cite{even2023grayc} and classify program validity as follows.
A program is \textbf{statically valid} if it conforms to the language specification and is expected to compile without errors.
A program is \textbf{dynamically valid} if it produces a well-defined, deterministic result at runtime, without triggering undefined, unspecified, or implementation-defined behaviours (\eg{} division by zero or array out-of-bounds accesses). Language-defined failures like exceptions or return code of \code{-1} are allowed if they conform to the language semantics.

Testing for miscompilation (silent errors during compilation that result in an incorrectly compiled program execution) is impossible with \textit{statically invalid} code, as such programs fail to compile entirely, producing no binary. Testing with \textit{dynamically invalid} programs can lead to false positives, where the observed issue is in the test program itself (\eg division by zero), rather than genuine compiler bugs.

In this tool paper, we propose a lightweight approach to enhance the static and dynamic validity test programs in fuzzing. Our new tool, \tool{}, seamlessly integrates into the LLM-based fuzzing workflow, demonstrated in the evaluation with black-, grey-, and white-box LLM-based fuzzing. The design of \tool utilises systematic feedback loop repair mechanisms, where compiler errors and sanitizer warnings guide iterative improvements of invalid test programs via LLM-suggested fixes of compiler and runtime issues, substantially increasing the proportion of valid test programs. \tool{} leverages locally deployed models as a safer alternative by preventing exposure of potentially sensitive code to external services, a crucial factor for industrial applications' developers.

We assess the validity rate of C programs and code coverage improvement on the LLVM/Clang compiler using black-, grey-, and white-box LLM-based fuzzers using various computer hardware. \tool significantly improved the static and dynamic validity rate and code coverage. Results show a strong dependency on hardware choice, with GPU-based configurations outperforming CPU-only setups. On GPU machines, \tool{} increased the validity rate from 47.0–49.4\% to 96.8–97.3\%, and improved code coverage in the LLVM/Clang compiler codebase by an absolute 0.3–21.2 percentage points across different compiler components.

\section{\tool{} Design}
\label{sec:desing}
\tool{} extends LLM-based fuzzing tools, enhancing the validity of LLM-generated test programs. It aims to address the gap between generating interesting test programs for testing compilers and the low validity rate common in LLM-based fuzzing. We refer to a program as valid if it is statically and dynamically valid. 

\para{Overview}
To improve the validity rate of LLM-generated test programs, we design a feedback-driven error-fixing loop using local LLMs. Failed attempts are retained for potential crash testing for testing efficiency. Importantly, \tool{} does not attempt to preserve the original program’s semantics or functionality during fixes, as doing so is unnecessary and would align more with code synthesis tasks than fuzzing.

\begin{figure}[t]
     \vspace{-0.2 cm}
     \centering
     \includegraphics[width=0.48\textwidth,height=0.09\textheight]{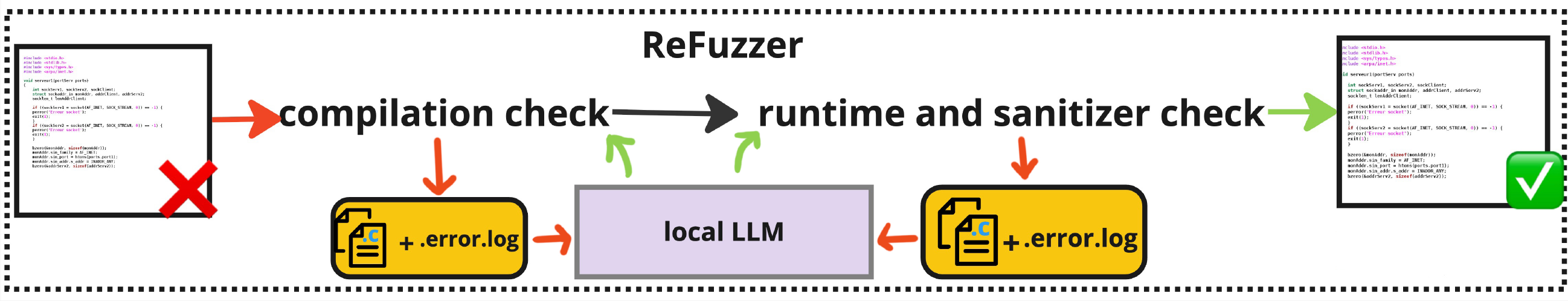}
     \vspace{-0.2 cm}
     \caption[\tool Workflow]{Overview of \tool.}
     \label{fig:refuzzer-workflow}
     \vspace{-0.5 cm}
\end{figure}

\autoref{fig:refuzzer-workflow} illustrates the systematic workflow of our approach: 
1) \tool{} takes C programs from LLM-based fuzzers like \white{} or \grey{} and checks whether they compile and pass sanitizer analysis.
2) Once a test program is detected to be statically or dynamically invalid,
\tool captures warnings and errors from the compiler or code sanitisers together with the C program and feeds it to a local LLM model using the following template:
\begin{tcolorbox}[colback=backgroundColour, colframe=black, boxrule=0.7pt, 
  before skip=5pt, after skip=5pt, left=2pt, right=2pt, top=2pt, bottom=2pt,
  title={Prompt template for fixing C test programs},
  fontupper=\footnotesize]
“Given the following C program and its compilation error log with \texttt{<<ARG1>>} optimisation level, analyze and correct the program to resolve \texttt{<<ARG2>>}.\textbackslash
n\texttt{<<PROGRAM-TO-FIX>>}”
\end{tcolorbox}
\noindent%
\code{ARG1} is a compilation level (\eg -O0) and \code{ARG2} is the error type ("compilation errors" or "sanitizer errors). Step 2 is repeated up to $n$ times until it succeeds in fixing all issues or the attempt limit is reached.
3) \tool, if succeeds, outputs a C program that is statically and dynamically valid.

\para{Feedback-Driven Error-Fixing Mechanism}
\tool{} utilises a feedback-driven approach as shown in \autoref{fig:refuzzer-workflow} between the test program code, the errors and warnings log (in yellow) and a local LLM (in grey-purple). \tool identifies invalid LLM-generated test programs by analysing compilation errors, warnings, and sanitiser outputs. 
These, along with the source code, are fed into a refinement loop powered by a local LLM. The loop suggests fixes, which \tool applies and verifies to produce valid test programs. Unfixable cases are moved to a crash-only folder and excluded from the seed bank to maintain quality. We use compilation output for static validity checks and code sanitizers for dynamic validity \cite{TSAN2009,asan2012,msan2015,ubsan_2017}.

Improving the validity rate of LLM-generated programs can lead to higher compiler code coverage, as valid programs are more likely to progress beyond the frontend and trigger middle- and backend optimisations, paths that invalid programs cannot reach. Example~\ref{ex:fix} demonstrates the result of refuzzing code.

\begin{example} \label{ex:fix}
\tool receives an invalid program shown in \autoref{fig:refuzzer-example:fixed:orig}. The original program contains assembly syntax errors and unsafe memory operations. The ASAN error.log: 

\noindent... \scode{==2352008==ERROR: AddressSanitizer: stack-buffer-overflow\\on address 0x7b4a700de02a at pc 0x5600114ff85b bp 0x7ffda0bb6f20\\sp 0x7ffda0bb66d0} ...

\noindent \tool{} fixes it and outputs program shown in \autoref{fig:refuzzer-example:fixed}, free of ASAN errors.

\tool{} 1) fixes errors in inline assembly instructions, 2) replaces non-standard functions with standard equivalents, 3) adds memory safety checks and corrects buffer allocations, and (4) eliminates potential dynamic errors reported by sanitizers.

\end{example}

\begin{figure}[!t]
    \centering
    \hspace*{0.15cm}
    \begin{minipage}{0.45\textwidth} 
        \centering
\begin{lstlisting}[basicstyle=\tiny, style=CStyle]
#include <stdio.h>
#include <string.h>
int main(int argc, char **argv) {
  // Use -O1 flag to enable optimization level 1
  __asm__("echo %1 > /proc/sys/kernel/osrelease\n", argv[0]);
  // Use deadargelim flag to eliminate dead arguments
  int num_args = sizeof(argv) / sizeof(argv[0]);
  for (int i = 0; i < num_args; i++) {
    if (!strcmp(argv[i], "-dead")) {
      argv[i] = NULL;
    }
  }
  // Use Collection of optimization passes
  int ret = vsprintf_s(NULL, "Hello, world! (%d args)\n", num_args);
  if (ret < 0) {
    printf("vsprintf_s failed: %d\n", ret);
  } else {
    printf("%s\n", argv[0]);
  }
  return 0;
}\end{lstlisting}
  \vspace{-0.3 cm}
  \caption{Original program generated via a black-box fuzzing calling \ollama{} \code{llama3.2}.}
  \label{fig:refuzzer-example:fixed:orig}
\end{minipage}
\hspace*{0.15cm}
\begin{minipage}{0.45\textwidth}
\centering
\begin{lstlisting}[basicstyle=\tiny, style=CStyle]
#include <stdio.h>
#include <string.h>
#include <stdarg.h>
#include <stdlib.h>
int main(int argc, char **argv) {
  // Ensure we have valid arguments
  if (argc < 1 || argv == NULL || argv[0] == NULL) {
    return 1;
  }
  // Use -O1 flag to enable optimization level 1
  __asm__("echo %0 > /proc/sys/kernel/osrelease\n" : : "r" (argv[0]));
  // Use deadargelim flag to eliminate dead arguments
  int num_args = argc;
  for (int i = 0; i < num_args && argv[i] != NULL; i++) {
    if (!strcmp(argv[i], "-dead")) {
      argv[i] = NULL;
    }
  }
  // Use Collection of optimization passes
  char buffer[512]; // Increased buffer size for safety
  int ret = vsprintf(buffer, "Hello, world! (%d args)\n", num_args);
  if (ret < 0) {
    printf("vsprintf failed: %d\n", ret);
  } else {
    printf("%s\n", buffer); // Print formatted buffer instead of argv[0]
    if (argv[0] != NULL) {
      printf("Program name: %s\n", argv[0]);
    }
  }
  return 0;
}\end{lstlisting}
 \vspace{-0.3 cm}
 \caption{Program fixed with \tool{}.}
 \label{fig:refuzzer-example:fixed}
\end{minipage}
\vspace{-0.5 cm}
\end{figure}

\para{Utilising Local LLMs}
We design a refinement loop using \textit{local} LLMs to fix invalid programs. Based on suggestions from the refinement loop, \tool{} applies targeted edits via an LLM and generates a corrected test program, which is then re-verified using the prompt template along with code analysis outputs from the refuzzed program. In this design, \tool{} executes models locally instead of cloud-based APIs, and can be configured to use any model available through \ollama{} \cite{ollama2025} and open-source platforms like Hugging Face~\cite{huggingface2025}, allowing the user to substitute any locally runnable language model of their choice. 

This flexibility allows users to balance computational requirements, LLM versions, and data privacy with their specific testing needs and hardware capabilities.
This design ensures reproducible evaluation, offers high reusability for users without access to paid platforms, and keeps all code, error logs, and corrections within the user's security perimeter--critical for testing compilers with proprietary or security-sensitive code.

\para{Implementation}
We implemented \tool{} \cite{refuzzer_artifact,refuzzer_dataset} in C++ and Python to automate multiple binary executions and generate a test suite. 
Our current implementation supports refuzzing of C and C++ programs.
We utilised \texttt{LLaMA 3.2} LLM due to its strong ability to interpret error messages and apply fixes in C and C++.
We set the number of refuzzing attempts to be two per test program. 

Nonetheless, as LLM-based approaches are programming language agnostic, our implementation supports test suites generated by fuzzers in any programming language. Yet, since \tool{} relies on dynamic analysis via code sanitizers, it requires appropriate dynamic analysis for the language of the test programs. Extending \tool{} to non-C/C++ compilers requires integrating equivalent dynamic analysis tools for those languages.

\section{Evaluation}
\label{sec:methd}
\label{sec:eval}
\subsection{Methodology}
We evaluate \tool{}'s quality and efficiency across three configurations, black-, grey-, and white-box LLM-based fuzzing \textit{with locally installed LLMs}, by asking:  

\begin{tcolorbox}[colback=gray!07, colframe=black,boxrule=0.7pt, before skip=5pt, after skip=5pt, left=2pt,
  right=2pt,
  top=2pt,
  bottom=2pt]
    \textbf{RQ1:} \textit{To what extent does \tool{} improve the validity rate of LLM-generated test programs by automatically detecting and fixing invalid cases through refuzzing?}
\end{tcolorbox}

\begin{tcolorbox}[colback=gray!07, colframe=black,boxrule=0.7pt, before skip=5pt, after skip=5pt, left=2pt,
  right=2pt,
  top=2pt,
  bottom=2pt]
    \textbf{RQ2:} \textit{What is the impact on validity rates when executed on CPU vs GPU architectures?}
\end{tcolorbox}



\begin{tcolorbox}[colback=gray!07, colframe=black,boxrule=0.7pt, before skip=5pt, after skip=5pt, left=2pt,
  right=2pt,
  top=2pt,
  bottom=2pt]
    \textbf{RQ3:} \textit{How effectively does \tool{} increase compiler code coverage beyond the frontend, particularly in the middle and back ends, across the three different configurations?}
\end{tcolorbox}

\para{Fuzzing Configurations Approaches}
We evaluate:

\textbf{(1) \black{}.}
Our LLM-based black-box fuzzer for generating C test programs using \ollama{}, leveraging general programming and optimisation keywords as in \cite{dakhama2023searchgem5}. 

\textbf{(2) Fuzz4All.} A grey-box compiler fuzzer~\cite{c5:Xia2024}. 

\textbf{(3) WhiteFox.} A white-box compiler fuzzer \cite{c4:Yang2023}.

\para{Hardware Configurations}
LLMs are highly dependent on GPU acceleration: Using more powerful GPU clusters generally yields faster refinements, higher throughput, and better test generation quality at the cost of increased resource demands. To quantify this trade-off, we evaluated:

\textbf{(i) CPU.}
Intel Xeon D-1548 (8 cores, 2.0 GHz, 64 GB RAM); no GPU acceleration.

\textbf{(ii) GPU(x1).}
Intel Xeon Silver 4114 (20 cores, 2.2 GHz, 192 GB RAM); 1 NVIDIA Tesla P100 (12 GB).

\textbf{(iii) GPU(x2).}
2 AMD EPYC 7542 (64 cores, 2.9 GHz, 512 GB RAM); 2 NVIDIA Tesla V100S (32 GB each).


\para{Experimental Procedure and Setup} 
We first generated a test suite of LLM-generated programs over 24 hours with \black{}, \grey{}, and \white{} fuzzers. This results in 3 sets of test programs, one for each fuzzzer. We stored them in persistent storage. We then refuzzed each set separately. \tool{} compiles each test program with Clang at \texttt{-O0} to collect a compilation log, applies fixes if needed, and re-evaluates the corrected programs using sanitizers to ensure dynamic validity. We then assess the quality of the test suites with and without \tool{} intervention.

We fuzzed LLVM Clang \code{21.0.0} using GCOV \code{11.4.0} with LCOV \code{2.3.1} for coverage analysis. Each 24-hour fuzzing run used a 60 s timeout and 16 GB memory limit per refuzzing. We configured \tool{} to use \ollama{} \cite{ollama2025} version \code{0.5.7} with \texttt{LLaMA 3.2} across all 3 configurations.




\begin{table*}[!t]
\caption{Validity rates, number of tests, and ReFuzzing time per test across configurations.}
\vspace{-0.2 cm}
\label{tab:merged-validity-time-tests}
\centering
\small
\footnotesize
\resizebox{0.99\textwidth}{!}{
\begin{tabular}{@{}l|cccc|cccc|cccc@{}}
\midrule
& \multicolumn{4}{c|}{\textbf{CPU}} & \multicolumn{4}{c|}{\textbf{GPU(x1)}} & \multicolumn{4}{c}{\textbf{GPU(x2)}} \\
\textbf{Fuzzer} & \textbf{B Valid} & \textbf{A Valid} & \textbf{\# Tests} & \textbf{Time/Test} & \textbf{B Valid} & \textbf{A Valid} & \textbf{\# Tests} & \textbf{Time/Test} & \textbf{B Valid} & \textbf{A Valid} & \textbf{\# Tests} & \textbf{Time/Test} \\
\midrule
\black & (46) 24.0\% & (155) 80.7\% & 192 & 14.2 s & (3821) 47.0\% & (7892) 96.8\% & 8150 & 4.9 s & (4103) 47.0\% & (8452) 96.8\% & 8732 & 3.1 s \\
\grey  & (85) 47.9\% & (198) 55.9\% & 285 & 15.1 s & (4042) 48.5\% & (8041) 96.6\% & 8321 & 4.6 s & (4320) 48.5\% & (8611) 96.6\% & 8911 & 2.9 s \\
\white & (25) 12.3\% & (112) 68.3\% & 204 & 14.5 s & (3847) 49.4\% & (7582) 97.3\% & 7791 & 5.2 s & (4117) 49.4\% & (8116) 97.3\% & 8341 & 3.5 s \\
\bottomrule
\end{tabular}
}
\vspace{-0.4 cm}
\end{table*}

\subsection{Results}

\para{RQ1 \& RQ2: Test Programs Validity Rate}%
For each fuzzer, \black, \grey, and \white{}, we measured the throughput and percentage of valid test programs with and without \tool{}. We then compared the improvement in validity against the additional time it required.

\autoref{tab:merged-validity-time-tests} shows the number of valid test programs before \tool (B column) and after (A column) applying \tool, the total number of generated test programs (\# Tests column), and the average processing time by \tool{} per test program (including up to one retry attempt when the initial fix fails; Time/Test column), across different hardware configurations (CPU, GPU(x1) and GPU(x2) column).

The results show a significant increase in valid test programs across all fuzzers after applying \tool{}, reaching a stable dynamic validity rate of 96.6-97.3\% with GPU setups, and slightly lower rates of 55.0-80.7\% on CPU-only configurations. On average, each test program took 14.2–15.1 s to be amended using a CPU-only setup, 4.6–5.2 s with GPU(x1), and 2.9–3.5 s with GPU(x2). While \tool{} significantly improves the validity rate, the time required for refuzzing is highly dependent on hardware. GPU-based setups offer a clear advantage, achieving results up to 4–5 times faster than CPU-only configurations. This emphasises the importance of hardware acceleration in LLM-based compiler fuzzing workflows.

\begin{tcolorbox}[colback=gray!20, colframe=black,boxrule=0.7pt, before skip=5pt, after skip=5pt, left=2pt,
  right=2pt,
  top=2pt,
  bottom=2pt]
    \textbf{RQ1 Answer. } Yes, \tool{} improved static \& dynamic validity rates to 96.6–97.3\%, with a stable processing time of 2.9–3.5 s per test on GPU(x2) across all three fuzzers, demonstrating that \tool{} operates efficiently with local LLMs.
\end{tcolorbox}

\begin{tcolorbox}[colback=gray!20, colframe=black,boxrule=0.7pt, before skip=5pt, after skip=3pt, left=2pt,
  right=2pt,
  top=2pt,
  bottom=2pt]
    \textbf{RQ2 Answer. } GPU(x2) and GPU(x1) achieved the same validity rate (96.6–97.3\%), with GPU(x1) requiring nearly twice as long. CPU-only execution was significantly longer with lower validity rates (55.0–80.7\%).
\end{tcolorbox}
\para{RQ3: Code Coverage}%
We analysed \tool{}'s impact on deep code coverage, focusing on execution paths that extend beyond the parser and frontend, which are likely unreachable when test programs contain compilation failures or are too simplistic. The results reported in RQ3 are based on the GPU(x2) configuration. Results for other hardware setups are available in GitHub \cite{refuzzer_cpu_data,refuzzer_gpu1_data} and show similar trends, but are omitted here due to page limit.

\autoref{tab:coverage} presents \textit{function coverage} measured via \code{lcov} and \code{gcov} instrumentation on the LLVM/Clang codebase. We evaluated function coverage percentage for LLM-generated test programs executed with the \black, \grey, and \white fuzzers, before (B column) and after (A column) refuzzing, with the coverage improvement presented in absolute percentage improvements ($\Delta$ column). The components in the table highlight improvements in the frontend (parsing and semantic analysis), intermediate representation (IR) generation, optimisation passes (e.g., loop optimisations, inlining, and dead code elimination) and backend code generation.

Our results show that \tool{} enhances function coverage, particularly in optimisations such as loop optimisation, inlining, and dead-code elimination (DCE). The highest improvement was observed with \black, achieving gains in inlining (+21.2\%), DCE (+17.0\%), and loop optimisation (+12.7\%). 
Followed by \white refuzzing, achieving significant improvements, particularly in inlining (+16.8\%) and DCE (+13.1\%). 
\tool{} had a lower yet meaningful impact on \grey, notably in inlining (+5.7\%) and DCE (+3.6\%). 
ven for \grey{}, the overall \textit{Opt. Passes} coverage increased from 0.2\% to 0.4\%  (doubling the coverage). This demonstrates that \tool{}’s improvements in test program validity enhance coverage beyond frontend components.

\begin{table}[t]
\caption{Function Coverage (\%) via \texttt{lcov} analysis. DCE: Dead Code Elimination. Analysis on GPU(x2) set, see full results in \cite{refuzzer_dataset}. 
(B) Raw coverage, (A) Enhanced coverage with \tool{}, $\Delta$: Absolute percentage improvement.}
\label{tab:coverage}
\vspace{-0.3 cm}
\footnotesize  
\raggedright 
\resizebox{0.47\textwidth}{!}{
\begin{tabular}{@{}lrrrrrrrrr@{}}
\toprule
\multirow{2}{*}{\textbf{Component}}                          & \multicolumn{3}{c}{\textbf{BlackBox}}           & \multicolumn{3}{c}{\textbf{Fuzz4All}}            & \multicolumn{3}{c}{\textbf{WhiteFox}}           \\ \cmidrule(l){2-10} 
                     & \multicolumn{1}{c}{\textbf{(B)}} & \multicolumn{1}{c}{\textbf{(A)}} & \multicolumn{1}{c}{\textbf{$\Delta$}} & \multicolumn{1}{c}{\textbf{(B)}} & \multicolumn{1}{c}{\textbf{(A)}} & \multicolumn{1}{c}{\textbf{$\Delta$}} & \multicolumn{1}{c}{\textbf{(B)}} & \multicolumn{1}{c}{\textbf{(A)}} & \multicolumn{1}{c}{\textbf{$\Delta$}} \\ \midrule
Frontend (Parser)    & 60.0 & 62.0 & +2.0     & 57.0 & 58.0 & +1.0     & 12.7 & 13.5 & +0.8     \\
AST \& Semantics     & 10.7 & 12.4 & +1.7     & 28.3 & 29.0 & +0.7     & 9.2  & 9.5  & +0.3     \\
IR Generation        & 33.0 & 43.2 & +10.2    & 9.2  & 9.6  & +0.4     & 7.2  & 7.6  & +0.4     \\
Opt. Passes          & 12.4 & 13.8 & +1.4     & 0.2  & 0.4  & +0.2     & 10.3 & 11.2 & +0.9     \\
•⁠  ⁠Loop Opt.          & 8.3  & 21.0 & +12.7    & 1.1  & 3.4  & +2.3     & 6.7  & 17.5 & +10.8    \\
•⁠  ⁠Vectorization      & 5.8  & 15.0 & +9.2     & 0.6  & 2.9  & +2.3     & 4.6  & 11.7 & +7.1     \\
•⁠  ⁠Inlining           & 11.8 & 33.0 & +21.2    & 0.3  & 6.0  & +5.7     & 9.2  & 26.0 & +16.8    \\
•⁠  ⁠DCE                & 17.0 & 34.0 & +17.0    & 0.2  & 3.8  & +3.6     & 13.4 & 26.5 & +13.1    \\
Backend Code Gen.    & 6.3  & 6.8  & +0.5     & 7.0  & 7.3  & +0.3     & 2.8  & 3.9  & +1.1     \\ 
                     \bottomrule
\end{tabular}
}
\vspace{-0.4 cm}
\end{table}

\begin{tcolorbox}[colback=gray!20, colframe=black,boxrule=0.7pt, before skip=5pt, after skip=10pt, left=2pt,
  right=2pt,
  top=2pt,
  bottom=2pt]
    \textbf{RQ3 Answer. } 
    Yes, \tool{} effectively increases compiler code coverage beyond the frontend, 
    most significantly in compiler optimisations, with improvements of up to +21.2\% in inlining, +17.0\% in dead code elimination, and +12.7\% in loop optimisation across the evaluated fuzzers.
\end{tcolorbox}

\section{Related Work \& Conclusion}
\label{sec:conclusion}
Code LLMs such as \texttt{StarCoder}~\cite{li2023starcoder} excel at code generation. Recent fuzzing approaches like \white~\cite{c4:Yang2023} and \grey{}~\cite{c5:Xia2024} leverage LLMs but depend on costly, closed-source models like \texttt{GPT-4}~\cite{achiam2023gpt} and often produce low-validity programs, as discussed in \cref{sec:intro}. Notably, \texttt{RoCode}~\cite{rocodeICSE2025} has explored improving code quality with a focus on static validity in general programming tasks, while our focus is on enhancing compiler code coverage in the context of fuzzing via both static and dynamic validity.

In future work, we plan to explore corpus and test program minimisation techniques aimed at maximising compiler code coverage while reducing refuzzing costs, an increasingly relevant objective in the context of LLM scalability and green computing. Further, we will explore the adoption of \tool{} to other programming languages, LLM-based tools, and fuzzing approaches. Building on the findings of RQ2, we will experiment with ChatGPT-style platforms on distributed GPU clusters to assess how backend architecture influences test validity and repair effectiveness compared to local LLM deployments.

\section{Tool Availability}%
\label{sec:availability}
A video demonstration showcasing \tool{} end-to-end is available at~\cite{refuzzer_demo_video}.
\tool{} with data and results, is freely accessible as a Zenodo record \cite{refuzzer_dataset} and GitHub \cite{refuzzer_artifact}.
\tool{} can be installed via README or a pre-built Docker \cite{refuzzer_docker}.
Greybox and whitebox fuzzing were extended to utilise open-source LLMs via PRs \cite{nmdis1999fuzz4all2024,nmdis1999whitefox2024}, with the WhiteFox PR already accepted and merged.

\bibliographystyle{IEEEtran}
\bibliography{main}
\end{document}